
\input phyzzx.tex
\tolerance=5000
\sequentialequations
\overfullrule=0pt
\nopubblock
\def\ick{\eqalign}
\twelvepoint
\def\sla{\raise.16ex\hbox{$/$}\kern-.57em}
\def\Dsl{\kern.2em\raise.16ex\hbox{$/$}\kern-.77em\hbox{$D$}}
\def\parsl{\sla\hbox{$\partial$}}
\def\Asl{\kern.2em\raise.16ex\hbox{$/$}\kern-.77em\hbox{$A$}}
\def\bsl{\sla\hbox{$b$}}
\def\eps{\epsilon}

\def\sla{\raise.16ex\hbox{$/$}\kern-.57em}
\def\Dsl{\kern.2em\raise.16ex\hbox{$/$}\kern-.77em\hbox{$D$}}
\def\parsl{\sla\hbox{$\partial$}}
\def\Asl{\kern.2em\raise.16ex\hbox{$/$}\kern-.77em\hbox{$A$}}
\def\Bsl{\kern.2em\raise.16ex\hbox{$/$}\kern-.77em\hbox{$B$}}
\def\square{\kern1pt\vbox{\hrule height 1.2pt\hbox{\vrule width 1.2pt\hskip 3pt
   \vbox{\vskip 6pt}\hskip 3pt\vrule width 0.6pt}\hrule height 0.6pt}\kern1pt}

\def\a{\alpha}

\gdef\journal#1, #2, #3, 1#4#5#6{               
    {\sl #1~}{\bf #2}, #3 (1#4#5#6)}            
\def\mpl{\journal Mod. Phys. Lett., }
\def\np{\journal Nucl. Phys., }

\def\pr{\journal Phys. Rev., }
\def\prd{\journal Phys. Rev. D, }

\REF\weinsus{S. Weinberg, \prd 19, 1277, 1979;
L. Susskind, \prd 20, 2619, 1979.}

\REF\bhl{W.A. Bardeen, C.T. Hill, M. Lindner, \prd 41, 1647, 1990.}

\REF\namjon{Y. Nambu and G. Jona-Lasinio, \pr 122, 345, 1961.}

\REF\chern{R. Jackiw and S. Templeton, \prd 23, 2291, 1981; J. Schonfield,
\np B185, 157, 1981; S. Deser, R. Jackiw and S. Templeton, \journal Ann.
Phys., 140, 372, 1982.}

\REF\fretow{D.Z. Freedman and P.K. Townsend, \np B177, 282, 1981.}

\REF\kalram{M. Kalb and P. Ramond, \prd 9, 2273, 1974.}

\REF\cresch{E. Cremmer and J. Scherk, \np B72, 117, 1974.}

\REF\hag{C.R. Hagen, \prd 19, 2367, 1979.}

\REF\lah{A. Lahiri, Los Alamos preprint LA-UT-92-3477 1992.}

\REF\bwf{M. Leblanc, R.B. MacKenzie, P.K. Panigrahi, R. Ray,
International Journal of Modern Physics, (1994), (in press).}

\REF\lmrt{M. Leblanc, D.G.C McKeon, A. Rebhan, T.N. Sherry, \mpl A6, 3359,
1991.}

\REF\mck{D.G.C. McKeon, \journal Can. J. Phys., 69, 1249, 1991.}

\line{\hfill UdeM-LPN-TH-94-194}
\line{\hfill CRM-2169}
\titlepage
\title{DYNAMICAL MASS GENERATION FOR NON-ABELIAN GAUGE FIELDS WITHOUT THE
HIGGS}
\author{Didier Caenepeel$^{(1)}$ and Martin Leblanc$^{(1,2)}$}

\vskip 20 pt

{\baselineskip=12 pt
\it
\centerline {\bf \it $\ ^{(1)}$Laboratoire de Physique Nucl\'eaire }
\centerline {\bf \it and}
\centerline {\bf \it $\ ^{(2)}$Centre de Recherches Math\'ematiques}
\vskip 4 pt
\centerline {\bf \it Universit\'e de Montr\'eal }
\centerline {\bf \it Case postale 6128, succ. centre-ville}
\centerline {\bf \it Montr\'eal, (Qc), Canada}
\centerline {\bf \it H3C-3J7 }}

\abstract{ We present an alternative to the Higgs mechanism to generate masses
for non-abelian gauge fields in (3+1)-dimensions. The initial Lagrangian is
composed of a fermion with current-current and dipole-dipole
type self-interactions
minimally coupled to non-abelian gauge fields. The mass generation occurs
upon the fermionic functional integration. We show that by fine-tuning the
coupling constants the effective theory contains massive non-abelian gauge
fields without any residual scalars or other degrees of freedom. }

\endpage

{\noindent {\bf I. Introduction.}}
\smallskip
The standard model has been widely accepted as {\it the} theory of electro-weak
interactions. It has successfully accounted for all experiments to date,
making it perhaps one of the greatest successes of modern
theoretical physics. However, apart from
the unknown value of the top quark mass, one of the present mysteries in the
standard model is the absence of the Higgs particle in present experiments.
This fundamental particle is a scalar, which relegates it as the only one in
this category. Perhaps, we should see these arguments as indications for
looking for alternatives to the Higgs mechanism for generating masses for the
elementary fermions and vector bosons.

Over the last twenty years, other mechanisms to
account for mass generation in the standard model have been proposed such as
technicolor theories ${}^{\weinsus}$, and dynamical symmetry breaking via a
top-quark condensate ${}^{\bhl}$
in analogy with BCS theory of superconductivity and the
Nambu-Jona-Lasinio mechanism in nuclear structure ${}^{\namjon}$. The latter
mechanism generates the Higgs particle (not as a fundamental particle)
and its consequences with a four-fermion interaction.

In (2+1)-dimensions, it is well-known
that the addition of a topological Chern-Simons term to the gauge field
kinetic part provides a mechanism for gauge fields mass generation without
spoiling gauge invariance ${}^{\chern}$. In a relatively similar spirit,
in (3+1)-dimensions, attempts to reproduce the Chern-Simons term involves
a product of field strength $\epsilon^{\mu\nu\alpha\beta}
F_{\mu\nu}\,F_{\alpha\beta}$. However, since this expression can be written
as a total derivative, it cannot bring any modifications to physics.
It is therefore necessary to introduce another field if we want to
mimic the Chern-Simons mechanism in (3+1)-dimensions. For instance, Freedman
and Townsend (F-T) and others
have developed theories containing an antisymmetric tensor
field coupled to an abelian gauge field ${}^{\fretow,\kalram,\cresch,\hag}$
$$
{\cal L} = -{1\over 8}H^\mu H_\mu -{1\over 4}F_{\mu\nu}^2
+{1\over 4} m \eps^{\mu\nu\alpha\beta}b_{\mu\nu}F_{\alpha\beta}
\eqn\tac
$$
where $H^\mu=\eps^{\mu\nu\alpha\beta}\partial_\beta b_{\nu\alpha}$,
$b_{\mu\nu}$
stands for an antisymmetric tensor field and $F_{\mu\nu}$ is the field strength
for the abelian gauge field. A generalization containing non-abelian
gauge fields
and non-abelian antisymmetric tensor fields is provided by F-T
$$
{\cal L}=
-{1\over 8} {\cal A}_\mu^a K^{\mu\nu, ab} {\cal A}_\nu^b
-{1\over 4} (F^a_{\mu\nu})^2
-{1\over 8} {e\over g} \eps^{\mu\nu\alpha\beta} b_{\mu\nu}^a F_{\alpha\beta}^a
\eqn\tic
$$
where ${\cal A}_\mu^a= (K_{\mu\nu}^{ab})^{-1} \eps^{\nu\alpha\beta\gamma}
\bigl (\partial_\gamma
b_{\alpha\beta}^b + e f^{bcd} A^c_\gamma b^d_{\alpha\beta}\bigr )$,
$K^{\mu\nu ab}\equiv g^{\mu\nu}\delta^{ab}- g f^{abc}\eps^{\mu\nu\rho\sigma}
b_{\rho\sigma}^c$ and $f^{abc}$ are the gauge group structure constants.
The theory \tac\  describes a massive photon of mass $m$ while the model \tic\
 is equivalent to the classical model of
massive non-abelian gauge fields of mass ${e\over 2 g}$ where
$e$ and $g$ are the coupling constants relative to gauge fields and
antisymmetric tensor fields respectively.
Both theories respect gauge invariance and have interesting vector gauge
symmetries.

Recently, a mechanism for photon mass generation in
(3+1)-dimensions has been suggested ${}^{\bwf}$,
which uses ideas present in the content of the abelian theory \tac .
The mechanism consists of a functional integration over fermions
minimally coupled to a low-energy abelian gauge field. The fermions
self-interacts via two types of contact interactions:
current-current and dipole-dipole terms. An antisymmetric tensor
field is introduced via the Hubbard-Stratonovich transformation to perform
the fermion's integration. After imposing conditions on the coupling
constants of the theory, it is possible to write the low-energy effective
action as the abelian model discussed in Ref.[\fretow,\kalram,\cresch,\hag],
which reproduces a massive abelian gauge field.

Since the massive mediators of forces present in the weak
interactions are known to be of non-abelian nature, it is perhaps
a good idea to reproduce the above argument generalized to the case of
non-abelian gauge
fields. Thus we begin in section II with the definition of our
model at the high-energy scale $\Lambda$. In section III, we integrate the
fermions up to energy $\Lambda$ where new physics could occur. In section IV,
we collect the radiative corrections, find the effective current at
low-energies and write the low-energy effective theory for the gauge and tensor
fields. In the appendix, we analyze the physical content of the
resulting gauge invariant theory.
\medskip
{\noindent {\bf II. High-energy model at the scale $\Lambda$.}}
\medskip

We begin with the following non-renormalizable but SU(N) gauge invariant
Lagrangian at the high-energy scale $\Lambda$
in which a fermion field is minimally
coupled to non-abelian gauge fields
$$
{\cal L}= -{1\over 2} \tr F_{\mu\nu}F^{\mu\nu}
          + \bar\psi(i\Dsl -m)\psi
          -g_1 \tr j_\mu j^\mu - g_2 \tr j_{\mu\nu} j^{\mu\nu}
\eqn\bozo
$$
where $D_\mu=\partial_\mu - ig A_\mu$ is the covariant derivative.
The non-abelian gauge fields are
defined by $A_{\mu }=A_{\mu}^a T^a$ where $T^a$ are the Lie algebra
generators obeying the commutation relations
$[T^a,T^b]=iC^{abc}T^c$ and the trace relation $\tr \bigl ( T^a T^b\bigr )
={1\over 2} \delta^{ab} $.

The last two quantities in \bozo\ are the current-current
and dipole-dipole fermionic self-interactions. The four-vector current and the
dipole current are given respectively
by $j_\mu~=~j_\mu^a~T^a$ and $j_{\mu\nu}=j_{\mu\nu}^a T^a$ with components
$$\ick {
j_\mu^a &= \bar\psi\gamma_{\mu}T^a\psi\cr
j_{\mu\nu}^a &= \bar\psi\sigma_{\mu\nu}\gamma_5 T^a\psi \cr }
\eqn\frank
$$
both of which transform in the adjoint representation of the SU(N) gauge
group, that is
$$\ick {
j_\mu^{a\prime} &= U (\theta) j_\mu^a U^{-1}(\theta) \cr
j_{\mu\nu}^{a\prime} &= U (\theta) j_{\mu\nu}^a U^{-1}(\theta) \cr}
\eqn\dick
$$
with the usual transformations on fermion field and gauge fields
$$
\ick {
\psi^\prime &= U (\theta) \psi \cr
A_\mu^\prime &= U (\theta) A_\mu U^{-1} (\theta) - {i\over g} \bigl (
\partial_\mu U (\theta) \bigr ) U^{-1} (\theta) \cr
F_{\mu\nu}^\prime & = U (\theta) F_{\mu\nu} U^{-1}(\theta) \cr }
\eqn\feyn
$$
It is easy to see that the Lagrangian
\bozo\ is invariant under the transformations \dick-\feyn\ .
The mass dimensions of the fields are $[\psi]=3/2$, $[A_\mu]=1$, and
$[g]=0$, $[g_1]=[g_2]=-2$ for the coupling constants.

In components, the Lagrangian \bozo\ can be written as
$$
\eqalign{
	{\cal L} =&-{1\over 4}F^a_{\mu\nu} F^{a,\mu\nu}
	+ \bar\psi(i\Dsl -m)\psi\cr
	&-{g_1\over 2}(\bar\psi\gamma^{\mu}T^a\psi)(\bar\psi\gamma_{\mu}T^a\psi)
	-{g_2\over 2}(\bar\psi\sigma^{\mu\nu}\gamma_5 T^a\psi)
       	(\bar\psi\sigma_{\mu\nu}\gamma_5 T^a\psi)\cr}
\eqn\a
$$
where the field strength is given by
$$
F_{\mu\nu}^a=\partial_{\mu}A_{\nu}^a-\partial_{\nu}A_{\mu}^a +
gC^{abc}A_{\mu}^b A_{\nu}^c
\eqn\toe
$$
This model is interesting because of the form of the dipole-dipole interaction,
which makes it different from the one studied in Ref.[\bhl].

{}From now on, for definiteness and due to obvious potential applications,
we will consider only the SU(2) gauge group with the usual su(2) Lie
algebra given by the Pauli matrices $T^a={\tau^a\over 2}$, $a=1,2,3$
and structure constants given by $C^{abc}=\epsilon^{abc}$.

We next apply the Hubbard-Stratonovich transformation by
introducing auxiliary non-abelian antisymmetric tensor
fields $b_{\mu\nu}$, which belong
also in the su(2)-Lie algebra, transform according to the adjoint
representation and have mass dimension $[b^a_{\mu\nu}]=1$.
Their introduction permit us
to rewrite the dipole-dipole interaction as
$$
-{g_2\over 2}(\bar\psi\sigma^{\mu\nu}\gamma_5 {\tau^a\over 2} \psi)^2\to
-{1\over 2g_2} b^a_{\mu\nu}b^{a \mu\nu} +
ib_{\mu\nu}^a(\bar\psi\sigma^{\mu\nu}\gamma_5 {\tau^a\over 2}\psi)
\eqn\goose
$$
since one can regain the original Lagrangian by solving the equation of motion
for $b_{\mu\nu}^a$ and substituting the result in the Lagrangian. As noted in
Ref.[\bwf],
we could also transform, in a similar way, the current-current interaction by
introducing other auxiliary gauge fields $a_{\mu}^a$. In what follows, we
choose instead to consider only the introduction of auxiliary antisymmetric
tensor fields and treat perturbatively the remaining four-fermion term.

\medskip
{\noindent {\bf III. Fermionic functional integration.}}
\medskip

We calculate the effective action for non-abelian
gauge fields and antisymmetric
tensor fields, which we treat as low-energy external fields,
by integrating out the fermions
$$
e^{i\Gamma_{\rm eff}[A^a,b^a]}=\int [{\cal D}\bar\psi][{\cal D}\psi]
	e^{i\int d^4x\left\{ -{1\over 4}F_{\mu\nu}^a F^{a\mu\nu}
	-{1\over 2g_2}b_{\mu\nu}^a b^{a\mu\nu}
	+\bar\psi(i\parsl -m+g\Asl^a{\tau^a\over 2}+\bsl^a{\tau^a\over 2})\psi
	-{g_1\over 2}(\bar\psi\gamma^{\mu}{\tau^a\over 2}\psi)^2\right\}}
\eqn\didier
$$
where $\bsl^a\equiv i \sigma^{\mu\nu}\gamma_5 b_{\mu\nu}^a$. The first
two terms, being external fields, can be put out of the functional integral.
The last term makes the integration difficult since it is not quadratic in the
fermionic fields.  Instead, we expand it in a power series in $g_1$
$$
e^{i\Gamma_{\rm eff}[A^a,b^a]}=e^{i\int d^4x\left(-{1\over 4}
	F_{\mu\nu}^aF^{a\mu\nu}-{1\over 2g_2}b_{\mu\nu}^a b^{a\mu\nu}\right)}
	\int [{\cal D}\bar\psi][{\cal D}\psi]e^{iS_0}
	\left(1-i{g_1\over 2}(\bar\psi\gamma^{\mu}{\tau^a\over 2}\psi)^2
	+\dots\right)
\eqn\martin
$$
where $S_0=\bar\psi(i\parsl -m+g\Asl^a{\tau^a\over 2}+ \bsl^a{\tau^a\over
2})\psi$ represents the bilinear fermionic action.
Using the approximation of vacuum domination we are able to rewrite the
current-current interaction in the form
$$
\eqalign{
	i{g_1\over 2}\int [{\cal D}\bar\psi][{\cal D}\psi]e^{iS_0}
        (\bar\psi\gamma^{\mu}{\tau^a\over 2}\psi)^2
        &=i{g_1\over 2}\left(\int [{\cal D}\bar\psi][{\cal D}\psi]e^{iS_0}
	\right)\langle 0|(\bar\psi\gamma^{\mu}{\tau^a\over 2}\psi)^2
	|0\rangle\cr
        &\simeq i{g_1\over 2}\left(\int [{\cal D}\bar\psi][{\cal
        D}\psi]e^{iS_0}\right)|\langle 0|(\bar\psi\gamma^{\mu}{\tau^a\over
        2}\psi)|0\rangle|^2\cr}
\eqn\d
$$
and the effective action can be rewritten to order $g_1$ by reconstructing
the exponential as
$$
e^{i\Gamma_{\rm eff}[A^a,b^a]}=e^{i\int d^4x\left(-{1\over 4}
        F_{\mu\nu}^aF^{a\mu\nu}-{1\over 2g_2}b_{\mu\nu}^a b^{a\mu\nu}
	-{g_1\over 2}\langle j^a_{\mu}\rangle^2 \right)}e^{i\Gamma^{(f)}_0}
\eqn\e
$$
where $\langle j^a_{\mu}\rangle$ is the current expectation value
and where the fermionic contribution
$$
\eqalign{
	\Gamma^{(f)}_0&=-i\Tr\log (i\parsl-m+g\Asl^a{\tau^a\over 2}
	+ \bsl^a{\tau^a\over 2})\cr
        &=\sum_{n=1}^\infty{i\over n}(-1)^n\Tr\left({1\over i\parsl-m}
	(g\Asl^a+\bsl^a){\tau^a\over 2}\right)^n\cr}
\eqn\f
$$
will be evaluated in a derivative expansion up to two derivatives on fields.
In the last equality the trace is taken over spinor indices, group indices
and momenta.

Let us start with a global explanation of our strategy. In contrast
to the abelian case where an effective theory for photons was derived only up
to two-point functions [that is $n=2$ in eq.\f], in the non-abelian effective
theory, we will need to go up to four-point functions. The reasons for this
will become clear in the course of the calculations. In part this is guided by
the presence of the SU(2) gauge invariance and also by
noticing that the self-coupling of the non-abelian gauge fields already
contains terms with four fields due to the nonlinearity of the theory.
Competition against those expressions will result in our model.

The contribution to $\Gamma_{\rm eff}$ quadratic in non-abelian
gauge fields and in antisymmetric tensor fields
is obtain when we put $n=2$ in the last expression
$$
{i\over 2}\tr\left({\tau^a\over 2}{\tau^b\over 2}\right)
\tr\int{d^4p\over(2\pi)^4}\bra{p}
{1\over i\parsl-m}(g\Asl^a+\bsl^a){1\over i\parsl-m}(g\Asl^b+\bsl^b)
\ket{p}.
\eqn\garfield
$$
We evaluate it using the cutoff parameter $\Lambda$ at which our model
\bozo\ is defined in order to
control the infinite logarithmic divergence. To ensure the SU(2) gauge
invariance with respect to $A_{\mu}$, we use a Pauli-Villars regularization
in conjunction with the cutoff regularization method
when computing its kinetic part.
Three different quantities are obtained~:
$$
\eqalign {
\Gamma^{(f)}_0[A^a,A^b]&=-{1\over 48\pi^2}g^2\delta^{ab}\log(\Lambda^2/m^2)
\int d^4x
A_{\mu}^a
\bigl ( g^{\mu\nu}(i\partial)^2-i\partial^{\mu}i\partial^{\nu}
\bigr )A_{\nu}^b,\cr
\Gamma^{(f)}_0[A^a,b^b]&={m\over 8\pi^2}g\delta^{ab}\log(\Lambda^2/m^2)
\int d^4x\quad
\epsilon^{\alpha\beta\mu\nu}b_{\alpha\beta}^b\partial_{\mu}A_{\nu}^a,
\cr
\Gamma^{(f)}_0[b^a,b^b] & ={1\over 8\pi^2}\delta^{ab}\log(\Lambda^2/m^2)
\int d^4x\left(m^2b^{a,\mu\nu}b_{\mu\nu}^b
+{1\over 6}b^{a\nu}_{\rho}[g^{\rho\mu}(i\partial)^2
-4(i\partial^{\mu})(i\partial^{\rho})]b_{\mu\nu}^b\right). \cr}
\eqn\bis
$$
Note that we have kept only the leading contributions
since we have assumed that $\Lambda\gg m$. The calculations
to this order do not differ from the abelian case (c.f. Ref.[\bwf])
except for the overall group theoretic factor
$ {\rm tr} \bigl ({\tau^a\over 2}{\tau^b\over 2}\bigr ).$

Next we compute the three and four-point functions involving only gauge
fields. This is done by putting $n=3$ and $n=4$ respectively.
Since gauge invariance
needs to be respected with respect to $A_{\mu}^a$, we continue to use a gauge
invariant regulator i.e. the Pauli-Villars method in conjunction with
the cutoff. The result then completes the first term of
Eq.~\bis\  to form the kinetic energy and self-coupling
$F_{\mu\nu}^a F^{a,\mu\nu}$ with its logarithmic divergence.

The rest of the calculation is performed with at least one antisymmetric
tensor field. We compute the three-point amplitudes. We
obtain three different contributions. The most interesting one is given by
$$
\Gamma^{(f)}_0[A^a,A^b,b^c]={m\over 16\pi^2}g^2\epsilon^{abc}
\log(\Lambda^2/m^2)\int
d^4x\epsilon^{\alpha\beta\mu\nu}A_{\mu}^aA_{\nu}^bb_{\alpha\beta}^c,
\eqn\boum
$$
because it completes the second expression of Eq.~\bis\  in order to make it
SU(2) gauge invariant. This is the desired $B\wedge F$-term.
The second less important but still present contribution is
$$
\eqalign{
\Gamma^{(f)}_0[A^a,b^b,b^c]&={i\over 48\pi^2}g\epsilon^{abc}
\log(\Lambda^2/m^2)\int d^4x\Bigl(
2A^{a\mu} [2b_{\mu\nu}^b (i\partial^{\alpha}) b_{\alpha}^{c\nu}
+(i\partial^{\alpha}) b_{\mu\nu}^b b_{\alpha}^{c\nu}]\cr
&+2A^{a\alpha} [2b_{\mu\nu}^b (i\partial^{\mu}) b_{\alpha}^{c\nu}
+(i\partial^{\mu}) b_{\mu\nu}^b b_{\alpha}^{c\nu}]
-A^{a\alpha} [b_{\mu\nu}^b (i\partial_{\alpha}) b^{c\mu\nu}
+(i\partial_{\alpha}) b_{\mu\nu}^b b^{c\mu\nu}]\Bigr) \cr}
\eqn\so
$$
The last three-point contribution contains three antisymmetric tensor
fields $\Gamma^{(f)}_0[b^a,b^b,b^c]$ and carries also a logarithmic
divergences. Fortunately, it is not necessary to compute it here since our
scaling argument (see below) will help us to remove it from the effective
theory. We request the reader to keep this in mind.

Finally we have to compute the four-point amplitudes by taking $n=4$ in Eq.~\f.
As stated before we have computed the quartic term in $A^a_{\mu}$.
The two expressions involving one or three $A^a_{\mu}$ fields
($\Gamma^{(f)}_0[A^a,b^b,b^c,b^d]$ and $\Gamma^{(f)}_0[A^a,A^b,A^c,b^d]$)
vanish since they involve a trace over
an odd number of gamma matrices. The quartic contribution
involving antisymmetric tensor fields does not need to be evaluated (again
see below).
Then, the only remaining quantity to be
computed is the one involving two $A^a_{\mu}$ and two $b^b_{\mu\nu}$ fields.
Evaluation of the leading ultraviolet divergence for
$\Gamma^{(f)}_0[A^a,A^b,b^c,b^d]$ gives
$$
\eqalign{\Gamma^{(f)}_0[A^a,A^b,b^c,b^d]={1\over 16\pi^2}g^2
& \Delta^{abcd} \log(\Lambda^2/m^2)  \cr
&\int d^4x \Bigl(2A^{a\mu} A^{b\nu} b_{\nu\alpha}^c b_{\mu}^{d\alpha}
+2 A^{a\mu} A^{b\nu} b_{\mu\alpha}^c b_{\nu}^{d\alpha}
-A^{a\mu} A_{\mu}^b b_{\nu\alpha}^c b^{d\nu\alpha} \Bigr)\cr}
\eqn\fou
$$
where $\Delta^{abcd} = \bigl [ \tr({\tau^a\over 2}{\tau^b\over 2}{\tau^c\over
2}
{\tau^d\over 2}) - {1\over 3} \delta^{ad}\delta^{bc}\bigr ] $, which is a
sum of products of two delta.

This completes the computation of the radiative corrections to one-loop order.

\medskip
{\noindent {\bf IV. Low-energy effective action.}}
\medskip

We are now in a position to evaluate the current, which will allow us to
find the first order correction in $g_1$ to the effective action.
{}From Eqs.~\bis-\fou, the effective current arising from the presence
of matter coupled to the non-abelian gauge fields is given by
$$
\eqalign{
\langle j^{a\lambda}\rangle={\delta\Gamma^{(f)}_0\over\delta A^a_{\lambda}}
=
&{m\over 8\pi^2}\log(\Lambda^2/m^2)\epsilon^{\lambda\mu\nu\alpha}
(\partial_{\alpha}b_{\mu\nu}^a+\epsilon^{abc}b_{\mu\nu}^b A^c_{\alpha})
 \cr
&+{i\over 48\pi^2}g\epsilon^{abc}\log(\Lambda^2/m^2)[2b_{\nu}^{b\lambda}
(i\partial_{\alpha}) b^{c\nu\alpha} +\dots] \cr
&+{1\over 8\pi^2}g^2\Delta^{abcd} \log(\Lambda^2/m^2)[2A^{b\nu} b_{\nu\alpha}^c
b^{d\lambda\alpha}+\dots ] \cr}
\eqn\jack
$$ where the ellipsis represents contributions to the current
of the same form as the term displayed in the corresponding square bracket but
with shuffled indices.
Note that the first expression above can be rewritten as
$$
{m\over 24\pi^2}\log(\Lambda^2/m^2)\epsilon^{\lambda\mu\nu\alpha}
H_{\alpha\mu\nu}^c
$$
where $H_{\alpha\mu\nu}^c=(D_{\alpha}b_{\mu\nu}+{\rm cyclic})^c$ and $D_\mu$
is the covariant derivative written in the adjoint representation
of the Lie algebra.

Now is time for serious recollection of the results. The effective current
enters the effective action to order $g_1$ upon squaring
Eq.~\jack. The current-current interaction therefore
contains a logarithmic divergence squared. In order, to
write an effective theory with antisymmetric tensor fields and
non-abelian gauge fields treated on the same foot,
and for which the kinetic term of the antisymmetric tensor fields has no scale
dependence at the tree level, we need to rescale $b_{\mu\nu}^a$ as
$$
\sqrt{g_1}{m\over 4\pi^2}\log(\Lambda^2/m^2)\; b_{\mu\nu}^a
\to b_{\mu\nu}^a
\eqn\pet
$$

Without any further assumptions, we obtain for the low-energy effective
Lagrangian density,
$$
\eqalign{
{\cal L}_{\rm eff}=
&-{1\over 4}F_{\mu\nu}^a F^{a\mu\nu}\left(1+{1\over 12\pi^2}
	g^2\log{\Lambda^2\over m^2}\right)
	+{1\over 12}H_{\mu\nu\rho}^a H^{a\mu\nu\rho}+{g\over 4\sqrt{g_1}}
	\epsilon^{\alpha\beta\mu\nu}b_{\alpha\beta}^a F_{\mu\nu}^a \cr
&+{1\over 2g_2}\left({16\pi^4\over g_1m^2\log(\Lambda^2/m^2)}\right)
	\left({m^2\over 4\pi^2}g_2\log{\Lambda^2\over m^2}-1\right) b_{\mu\nu}^a
	b^{a\mu\nu} \cr
&-{1\over 3}\left( {\pi^2\over g_1m^2\log(\Lambda^2/m^2)}\right)
	b^{a\nu}_{\rho}[g^{\rho\mu}\partial^2 -4\partial^{\mu}\partial^{\rho}]
	b_{\mu\nu}^a \cr
&-{1\over 3}\left( {g\pi^2\over g_1m^2\log(\Lambda^2/m^2)}\right)\epsilon^{abc}
	\left\{2A^{a\mu} [2b_{\mu\nu}^b (\partial^{\alpha} b_{\alpha}^{c\nu})
	+(\partial^{\alpha} b_{\mu\nu}^b) b_{\alpha}^{c\nu}]+\dots \right\} \cr
&+{1\over 6}\left({g\pi^2\over\sqrt{g_1}m^2\log(\Lambda^2/m^2)}\right)
	\epsilon^{\lambda\mu\nu\alpha}\epsilon^{abc}(D_\alpha b_{\mu\nu}^a)
	[2b_{\lambda\rho}^b  (\partial^{\sigma} b_{\sigma}^{c\rho}) +\dots ] \cr
&-\left({g^2\pi^2\over\sqrt{g_1}m^2\log(\Lambda^2/m^2)}\right)
	\Delta^{abcd}\epsilon^{\lambda\mu\nu\alpha}(D_\alpha b_{\mu\nu}^a)
	[2A^{b\rho} b_{\rho\sigma}^c b_{\lambda}^{d\sigma} +\dots ]
 \cr }
\eqn\gom
$$
where the first and third to the sixth term in Eq.~\gom\  were obtained by
direct computations. The second and the last two are instead obtained by
squaring the effective current. Incidentally, we left uncomputed the $b^3$ and
$b^4$ expressions because these are
suppressed by more than one logarithm of the
scale $\Lambda$. Eq.~\gom\  corresponds to the
effective action we wanted to order four in fields.

We now proceed with approximations that will help us to recover
the claims stated in the abstract.
In order to eliminate the mass term for the antisymmetric tensor fields,
we impose that the cut-off $\Lambda^2$ satisfy the gap equation
$$
m^2=\Lambda^2 e^{-4\pi^2/m^2g_2} \eqn\bof
$$ which goes within the perturbative expansion argument that the coupling
$g_2$ is small for sufficiently large $\Lambda$ and fixed fermion mass.
Upon using the gap equation, we next assume that the ratios
$$
{\pi^2\over g_1m^2\log(\Lambda^2/m^2)}={g_2\over 4g_1}
\eqn\top
$$
and
$$
{\pi^2g\over \sqrt{g_1}m^2\log(\Lambda^2/m^2)}={gg_2\over 4\sqrt{g_1}}
\eqn\tip
$$
are small, which permits us to smoothly get ride of the unwanted remaining
expressions in Eq.~\gom. The second ratio \tip\ is a necessary extra condition
present only in the non-abelian theory.

With the assumptions given above, and upon renormalizing the coupling
constant $g$ by absorbing the logarithmic divergence present in the
first quantity, we obtain a low-energy
effective Lagrangian of the form
$$
{\cal L}_{\rm eff}=
-{1\over 4}F_{\mu\nu}^a F^{a\mu\nu}
+{1\over 12}H_{\mu\nu\rho}^a H^{a\mu\nu\rho}
+{1\over 4\sqrt{g_1}}g
\epsilon^{\alpha\beta\mu\nu}b_{\alpha\beta}^a F_{\mu\nu}^a
\eqn\rip
$$
valid up to energy $m$ and
which describes massive SU(2) non-abelian gauge fields with
mass ${g\over {\sqrt g_1}}$ [see appendix].

\medskip
{\noindent {\bf V. Conclusions.}}
\medskip
We have succeeded in functionally integrating the four-Fermi theory to end up
with an effective $b\wedge F$-theory in agreement with the model proposed by
Lahiri ${}^{\lah}$ but different than the one proposed by Freedman-Townsend.
It is interesting to note that we did not reproduce Freedman-Townsend's model
\tic\ because
the nonlinearities in the antisymmetric tensor fields are suppressed by
the cutoff $\Lambda$. This is perhaps unfortunate
because the F-T model has higher reducible vector gauge invariance and behaves
properly for renormalization purpose ${}^{\lmrt}$. However, it has been shown
to have unitarity problems in tree-level scattering~${}^{\mck}$.

In the appendix, we have performed an analysis
to count the degrees of freedom of the theory \rip\ .
We have shown that it does
indeed contain, for each color,
two massive transverse modes
combined with a massive longitudinal mode of the same mass, which is
necessarily interpreted as the third degree of freedom for the
non-abelian gauge fields.

Of course, our presentation of the counting
of degrees of freedom here is restricted to the study of the
Lagrangian's bilinear part
and more has to be said in order to understand the interactions
of the model in this basis. Renormalizability and unitarity of
model \rip\ deserve an other study and generalization to include $SU(2)_L\times
U(1)_Y$ are presently under investigation.

\medskip
{\noindent {\bf Acknowledgements.}}
\medskip

We gratefully acknowledge D.~London, R.B.~MacKenzie
and M.B.~Paranjape for useful discussions and the
members of the SPD for their interest in the project.
This work was supported in part by the Natural Science and
Engineering Research Council of Canada and the Fonds F.C.A.R. du Qu\'ebec.
\vfill\eject
\medskip
{\noindent {\bf Appendix: Physical Content.}}
\medskip

It remains to show that the model in Eq.~\rip\  has the proper
number of degrees of freedom necessary to insure a description of
massive non-abelian gauge fields.  We follow Ref.[\hag] and absorb the gauge
field coupling constant into the gauge fields for convenience.
It is well-known that the pure gauge field part of the
the Lagrangian density \rip\  can be written in terms of the
canonical variable ${\bf A}^a$ and the electric
field (its conjugate)  ${\bf E}^a$, in the
following way
$$\eqalign {
{\cal L}_{F^2}&= -{1\over g^2} \Bigl \{
\partial_0 {\bf A}^a \cdot {\bf E}^a + {1\over 2}({\bf E}^a\cdot
{\bf E}^a + {\bf B}^a\cdot {\bf B}^a ) - A_0^a\bigl(
({\bf D}\cdot{\bf E})^a\bigr ) \Bigr \} \cr }
\eqn\gag
$$
where the electric and magnetic fields are
$$\ick {
{\bf E}^a &= F^{i0}_a=-\nabla A_0^a-\partial_0{\bf A}^a+\epsilon^{abc}
{\bf A}^b A_0^c \cr
{\bf B}^a &= -{1\over 2} \epsilon^{ijk} F^{jk}_a = \nabla\times{\bf A}^a
-{1\over 2}\epsilon^{abc}{\bf A}^b\times{\bf A}^c \cr }
\eqn\dip
$$
and the magnetic field can be read as a function of the variable ${\bf A}^a$.
In this way, $A_0^a$ is then easily identified as the Lagrange multiplier.

Upon integration by parts, the second term of the Lagrangian \rip\ can be
written as
$$\ick {
{\cal L}_{bF}&={1\over {\sqrt g_1}}v^{a,\mu} A^a_\mu +
{1\over 2{\sqrt g_1}}\epsilon^{abc}\bigl [
2 A_0^a \;{\bf a}^b\cdot{\bf A}^c - {\bf b}^a\cdot({\bf A}^b\times{\bf A}^c)
\bigr ]\cr}
\eqn\leb
$$
and the kinetic energy (and three and four bosons couplings)
for the antisymmetric tensor fields as
$$\ick {
&{\cal L}_{H^2} = {1\over 2}v^\mu_a v^a_\mu
+\epsilon^{abc}\bigl[ A_0^a{\bf v}^b\cdot{\bf a}^c - v_0^a{\bf A}^b\cdot
{\bf a}^c + ({\bf v}^a\times{\bf A}^b)\cdot{\bf b}^c\bigr] \cr
&-{1\over 4}\bigl [ A_\mu^a A^{a,\mu} b^c_{\nu\tau} b^{\nu\tau}_c -
A_\mu^a A^{c,\mu} b^a_{\nu\tau} b^{\nu\tau}_c\bigr ]
+{1\over 2}\bigl [ A_\mu^a A^a_\nu b^{c,\nu}_{\tau} b^{\mu\tau}_c -
A_\mu^a A^c_\nu b^{a,\nu}_\tau b^{\mu\tau}_c \bigr ] \cr}
\eqn\ban
$$
where we introduced the convenient notation
$$\ick {
v^\beta_a &= {1\over 2} \epsilon^{\mu\nu\alpha\beta}\partial_\alpha
b_{\mu\nu}^a \cr
{\bf a}^a &= {1\over 2}\epsilon^{ijk}b^a_{jk} \cr
{\bf b}^a &= -b_{0i}^a \cr }
\eqn\jop
$$

The dynamical content of the theory is located in the bilinear part in fields.
In order to exhibit the physical content of
the theory, we use the non-abelian gauge invariance
of the theory together with a change of variable motivated by
the vector gauge invariance of the Lagrangian \rip\ up to the bilinear part in
fields.
In order to proceed, we define vectors in space by the known decomposition into
longitudinal and transverse part: ${\bf J}={\bf J}_T+
{\bf J}_L$ such that $\nabla\cdot{\bf J}_T=0$ and $\nabla\times {\bf J}_L=0$.
In a non-abelian theory such as the present case, it is always possible to
choose for a fixed vector $n_\mu$, non-abelian gauge fields $n^\mu A^a_\mu=0$.
We choose this gauge fixing condition such that ${\bf A}_L=0$. In this gauge,
the Lagrangian's bilinear part in fields is
$$
{\cal L}^{\rm bilinear}=-{1\over 2g^2}\Bigl [
{\bf A}^a_T \cdot\square
{\bf A}^a_T + A_0^a\nabla^2A_0^a \Bigr ]
+{1\over{\sqrt g_1}}\Bigl [v^a_0 A^a_0 - {\bf v}^a\cdot{\bf A}^a_T\Bigr]
+{1\over 2}v^a_0 v^a_0 - {1\over 2}{\bf v}^a\cdot{\bf v}^a
\eqn\soo
$$where $v_0^a=-\nabla\cdot{\bf a}^a$ and ${\bf v}^a=\nabla\times{\bf b}^a
+\partial_0{\bf a}^a$.

Next, we note that the above Lagrangian's bilinear part in fields
is invariant under the vector gauge invariance
\def\bb {{\bf {\tilde b}}}
\def\aa {{\tilde {\bf a}}}
$$\ick {
{\tilde b}_{\mu\nu}^a&=b_{\mu\nu}^a+\partial_\mu\Gamma^a_\nu
-\partial_\nu\Gamma^a_\mu \cr}
\eqn\al
$$ without changing the non-abelian gauge fields.
Eq.~\al\  is not an invariance of the full Lagrangian. Perhaps
this is unfortunate, but we can still use a field redefinition to uncover the
physical degrees of freedom of the theory at the expense of introducing
complications in the interactions. For instance, we can always redefine the
antisymmetric tensor fields such that
$$\ick {
{\aa}^a &={\bf a}_L^a \cr
{\bb}^a  &=({\bf b}^a)_T - \partial_0{\bf \Gamma}^a_T\cr }
\eqn\ko
$$
that is, the new vector field ${\aa}^a$ has no transverse part and the
${{\bb}}^a$-field has no longitudinal part. Indeed, it is sufficient to
take
$$\ick{
{\bf \Gamma}_T^a &= -{\nabla\times {\bf a}^a_T \over \nabla^2}\cr
\Gamma_0^a &= {\nabla \cdot ({\bf b}^a)_L\over \nabla^2}\cr}
\eqn\lo
$$
Equipped with this
field redefinition, we substitute in the Lagrangian
$$\ick {
{\bf a}^a &\rightarrow {\aa}^a + \nabla\times{\bf \Gamma}^a_T\cr
{\bf b}^a & \rightarrow  {\bb}^a
+\partial_0{\bf \Gamma}^a_T - \nabla {\Gamma}_0^a \cr }
\eqn\lo
$$ with $\aa^a$ and $\bb^a$ given by eq.\ko, to obtain
$$\ick {
{\cal L}^{\rm bilinear}=&-{1\over 2g^2}\Bigl [
{\bf A}^a_T\cdot \square {\bf A}^a_T + A_0^a\nabla^2A_0^a \Bigr ]\cr
&-{1\over{\sqrt g_1}}\Bigl [ A^a_0\nabla\cdot{\aa}^a
+(\nabla\times{\bb}^a+\partial_0{\aa}^a)\cdot{\bf A}^a_T\Bigr]\cr
&-{1\over 2}{\aa}^a \cdot \nabla^2{\aa}^a
-{1\over 2}(\nabla \times \bb^a+\partial_0{\aa}^a)\cdot
(\nabla\times \bb^a+\partial_0{\aa}^a)\cr}
\eqn\ji
$$
showing clearly that the fields $A_0^a$ and $\bb^a$ do not propagate, as it
is the case for ${\bf a}^a_T$ and ${\bf b}^a_L$.

We eliminate the fields $A_0^a$ and $\bb^a$
by using their equation of motion to the order
considered $A_0^a=-{g^2\over {\sqrt g_1}\nabla^2}\nabla\cdot{\aa}^a$ and
$\bb^a= -{1\over {\sqrt g_1}}{\nabla\times{\bf A}_T^a\over \nabla^2}$
and substitute these into the bilinear part of the Lagrangian \ji, we get
$$\ick {
{\cal L}^{\rm bilinear} =& -{1\over 2g^2}
{\bf A}^a_T\cdot \bigl( \square - {g^2\over g_1}\bigr) {\bf A}^a_T
+ {1\over 2} {\aa}^a \cdot \bigl (\square - {g^2\over g_1}\bigr) {\aa}^a
\cr }
\eqn\ti
$$
exhibiting in this way the physical content of the theory, that is for each
group direction, two transverse modes with mass ${g\over {\sqrt g_1}}$ and one
longitudinal mode with the same mass. Since the transverse modes have gained
a mass from its coupling to the antisymmetric tensor fields, we are led
to interpret the longitudinal mode ${\bf a}_L^a$ as the third degree of
freedom for each massive non-abelian gauge fields, $a=1,2,3$,
hence the Lagrangian
\rip\ describes massive non-abelian gauge fields without any residual
degrees of freedom.

Of course,
the interactions have been modified by this choice of variables
but writing the interactions in this basis would not be
enlightening, so we do not pursue this further since we have achieved the
desired result of computing the number of degrees of freedom.

\refout
\bye